                     \documentclass[journal, final]{IEEEtran}
                     \usepackage{ifpdf}

                     \ifCLASSOPTIONcompsoc
                     \else
                       \usepackage{cite}
                     \fi
                     \usepackage{graphicx}
                     \DeclareGraphicsExtensions{.eps2}
                     \ifCLASSINFOpdf
       
                     \else
                       \DeclareGraphicsExtensions{.eps2}
                     \fi
                     \usepackage[cmex10]{amsmath}
                     \usepackage{algorithmic}

                     \usepackage{array}

                     \usepackage[tight,footnotesize]{subfigure}

                       \usepackage[caption=false,font=footnotesize]{subfig}
                     \usepackage{fixltx2e}

                     \usepackage{stfloats}

                     \usepackage{verbatim}
                     \usepackage{url}

\begin{document}
                     %
                     \title{A Survey of System Security in Contactless Electronic Passports}
                     %
                     %
                     %
                     %
                     
                     \author{Anshuman~Sinha,~\IEEEmembership{Engineering Lead, United Technologies Corporation, Fire \& Security\\anshuman.sinha2@fs.utc.com 
                     }}

                     \IEEEcompsoctitleabstractindextext{
                     \textbf{A traditional paper-based passport contains a Machine-Readable Zone (MRZ) and a Visual Inspection Zone (VIZ). The MRZ has two lines of the holder's personal data, some document data, and verification characters encoded using the Optical Character Recognition font B (OCR-B). The encoded data includes the holder's name, date of birth, and other identifying information for the holder or the document. The VIZ contains the holder's photo and signature, usually on the data page. However, the MRZ and VIZ can be easily duplicated with normal document reproduction technology to produce a fake passport which can pass traditional verification. Neither of these features actively verify the holder's identity; nor do they bind the holder's identity to the document. A passport also contains pages for stamps of visas and of country entry and exit dates, which can be easily altered to produce fake permissions and travel records. The electronic passport, supporting authentication using secure credentials on a tamper-resistant chip, is an attempt to improve on the security of the paper-based passport at minimum cost. This paper surveys the security mechanisms built into the first-generation of authentication mechanisms and compares them with second-generation passports. It analyzes and describes the cryptographic protocols used in Basic Access Control (BAC) and Extended Access Control (EAC).} 
                     
                     
                     }

                     \maketitle

                     \IEEEdisplaynotcompsoctitleabstractindextext

                     %
                     \IEEEpeerreviewmaketitle

                     %
                       \section{Introduction}\label{sec:introduction}\par
                     %

                     %
                     %
                     %
                     %
                     \IEEEPARstart{I}{n} an effort to secure the borders of the United States of America, Congress has legislated requirements for Electronic Passports (ePassports) \cite{14} for all visitors from countries participating in the Visa Waiver Program (VWP) \cite{DHSPrograms}. Any passport issued by a participating state after October 2010  must be machine-readable with an electronic facial image encoded on a secure chip. As a requirement of the US-VISIT program \cite{DHSVisit}, all visitors to the U.S. must have their photo, and a fingerprint of their index finger, taken for electronic comparison. In a reciprocal move, the U.S. has started issuing electronic passports to its citizens from all domestic issuance agencies since August 2007. 
                     
                     European Union countries have advanced to a second-generation electronic passport. Countries like Germany, France, and the Czech Republic have passports at the next level, whereby biometric information is stored in a secure passport chip. Similarly, Asian countries like Malaysia and European Union (EU) countries have already advanced to biometrics and Extended Access Control (EAC) like access control. Any effort to make passports electronic and secure requires adding hardware, firmware, and software at different levels to the existing verification infrastructure. The centerpiece of the next generation of passport technology lies in Java programmable secure controllers with advanced cryptographic capabilities.  An ePassport has an embedded Radio Frequency Identification chip (RFID) with processing capability for cryptographic computations. Unfortunately, the wireless link between this passport tag, which is called chip in the remainder of this paper, and a passport verification reader can lead to security and privacy threats. Threat modeling is based on first principles which analyzes compromise to common security goals of any system.

              \section{Threat Model for ePassports}
              Electronic passports \cite{Juels} must prevent the known attacks common to secure radio frequency-based identification and access control systems. Comprehensive threat modeling and designing countermeasures are key to the robustness of any security system; ePassports are no different in this regard. Each threat must have one or more mechanisms (or countermeasures) built into the system. A security system can be designed only to secure against known threats and attacks.  A threat is defined as a weak link in the system which can be broken to compromise system's security.  A threat model should be comprehensive enough to include all known attacks.  Threat and attack model will have overlaps, however this may not be the case for every noted threat.  
          
              \begin{table}[h]
              \renewcommand{\arraystretch}{1.3}
              \caption{ePassport Threat Model}
              \centering
              \begin{tabular}{ l l l } 
              \hline 
              S.No. & Security Threats & Compromises \\ [0.5ex] 
              \hline\hline 
              1. & Forging & Integrity \\
              2. & Repudiation & Availability \\
              3. & Skimming & Confidentiality \\
              4. & Cryptographic Analysis & Confidentiality \\
               & & \& Integrity \\
              5. & Unauthorized System & Authorization \\
              6. & Unauthorized User &  Verification \\
              7. & Privacy Threats & Consent, verification \\
              & & \& authentication \\
              8. & Platform Integrity &  Integrity \\
              \hline
              \end{tabular}
              \label{table:threatmodel}
              \end{table}
       
       \begin{twocolumn}
       
       However, system may fail to secure against unknown threats or those discovered after the design has been committed and the product taken to field, side channel attacks \cite{Kocher} is example of one of the threat which was not modeled and accounted for in the design of early hardware security systems.  Cryptographer these days are aiming to build systems which would be safe to unknown attacks.  A threat model for ePassports is given in Table~\ref{table:threatmodel}.  The left column shows known threats that must have``antidotes'' in the design to ensure that fundamental security requirements are not violated. The right column shows the requirement(s) at risk. System security, including passport system security, comprises the following fundamental security requirements: Secrecy - preventing system information from flowing in unauthorized paths; Integrity - preventing unauthorized modification of system state; Legitimate Use - preventing unauthorized use of system resources; Availability - preventing unauthorized interference in the use of the system. Notice that all of these issues involve the notion of authorization.
            
              	\paragraph{Forging} A passport can be forged by replacing a complete chip with a different one. There are two cases of such an attack. The first is replacement of a chip with a cloned or duplicate LDS, the unprotected contents of an ePassport's Logical Data Structure. In this case, the duplicated passport LDS matches that of an original electronic passport. This is a cloning of a chip and its complete data set. The second case is replacement of a chip with a tampered LDS. (A secure chip generally has hardware mechanisms built-in to protect against data alteration.) Forging clearly violates integrity.
       
              	\paragraph{Non-Repudiation} A non-repudiation attack concerns an ePassport that has been tampered with to withhold information embedded in its secure chip. The chip or its antenna could be set to a state where it can not be read by a valid reader. In such a scenario, the passport holder can not be verified electronically. Due to the non-availability of the secure chip, two situations emerge. The first is the failure to electronically authenticate the passport. If the secure chip is not available, the passport can not be authenticated and verified electronically. The authentication must be performed by relying on MRZ data. The second case that emerges is failure to verify the passport holder. Due to the non-availability of the passport chip, the holder's biometrics and other data stored on the chip may not be verified. If the passport holder's biometrics are stored only on the secure chip and not in a backend system, the biometrics cannot be verified. 
              
              	\paragraph{Skimming} The unprotected contents of an ePassport's LDS could be read by an unauthorized reader close enough to the passport to use the wireless link. A skimmer could utilize a reader which has been modified to read data from distances greater than the passport designers anticipated.
       
              	\paragraph{Cryptographic Analysis} The cryptographic keys stored in an ePassport chip could be exposed by deploying mass computing power after gathering skimmed or eavesdropped data from different electronic passports. The keys could be used to illicitly communicate with other passports.
       
       	\paragraph{Unauthorized System} A system may be authentic but may not have the correct privileges to read information from the passport.
       
       	\paragraph{Unauthorized User} Unauthorized user is anyone who is not the person to whom the identity credentials had been issued.  Such a person misuses the identity documents by using someone else's information or appearance.
       
       	\paragraph{Privacy Threats} Privacy threats \cite{PIVPrivacy} are unapproved use of personal information or tracking of the passport holder.
       
       	\paragraph{Platform Integrity} The platform on which applications run must be free from any malicious code which can act as a Trojan Horse for the information stored in the ePassport. Usually, such methods are added to the system for the ease of testing during development and must be removed to completely secure the system.
       
              \section{Known Attacks for ePassports}
              Security mechanisms built into ePassports must counter the specific known attacks shown in Table~\ref{table:risks}. The attack profile is as important as the threat profile.  
       
              \begin{table}[h]
              \renewcommand{\arraystretch}{1.3}
              \caption{ePassport Attack Model}
              \centering
              \begin{tabular}{ l l } 
              \hline 
              S.No. & Known Attacks \\ [0.5ex] 
              \hline\hline 
              1. & Forging  \\
              2. & Skimming  \\
              3. & Eavesdropping\\
              4. & Illicit Verification \\
              5. &  Data/Noise Injection  \\
              6. & Impostor/False Biometrics \\
              7. & Rogue Reader/Hardware  \\
              8. & Duplication/Cloning \\
              9. & Tracking \\
              \hline
              \end{tabular}
              \label{table:risks}
              \end{table}
              
       	\paragraph{Forging} A passport can be forged by replacing its complete chip with a different one. There are two such cases. First, the chip is replaced with a cloned or duplicated LDS. In this case, the chip on the duplicated passport matches the data contents of the original ePassport. This is cloning of a chip and its complete data set. Second, the chip is replaced with a tampered LDS. A new chip, with a modified copy of credentials is put in place of the original chip. Such a modified secure chip is embedded into a passport from which the original chip was removed. A secure chip has hardware mechanisms built to protect against data alteration.
       
       	\paragraph{Skimming} An ePassport can be skimmed to read the unprotected contents of its LDS. The skimmer may gather sensitive details like the passport holder's name, age, address, and travel information.  The skimmer could utilize readers which are modified, extended, or rogue to read data from distances greater than designed for.
       
              	\paragraph{Eavesdropping} Eavesdropping is an attack to intercept communication between card and reader. The stolen information could either be document or personal information or system related cryptographic information which can be replayed to illicitly communicate with other passports.
       
              	\paragraph{Illicit Verification} Illicit verifiers are like fake ATM(s) installed at locations to falsely retrieve the PIN of a banking card. The personal information could be skimmed if such a system with copied keys is installed. 
              
              	\paragraph{Data/Noise  Injection} A data injector is a device which can manipulate or alter the data being sent by a secure chip to a reader or vice versa. It can interject data with its own data frames which may not only interfere with communication but also alter add or subtract data exchanged between secure reader and chip.
              
              	\paragraph{Imposter/False Biometrics} An impostor can fake the biometrics of an authentic passport holder with methods like wax fingers or face masks to fool the system. Impostors break the security of a  system by faking the identity of the actual passport holder.    
       
       	\paragraph{Rogue Reader/Hardware} A modified reader can be used to read the contactless data from a card beyond it normal operating range. Other modified hardware can store communications while legitimately communicating with a reader.

       	\paragraph{Duplication/Cloning} Cloning of a chip and duplication of its complete or partial data is an attack requiring sophisticated machinery and means. Such attacks are conducted only by advanced attackers who have the finances to invest in such systems.
              
              	\paragraph{Tracking} A tracker is a person who is only interested in knowing where the passport holder is traveling. A tracker may not have access to all the information which is stored in the secure chip. The tracker is only interested to know the whereabouts of a person. The tracker skims the data which could be used to trace a passport holder's location. 
       
                     \section {Authentication in ePassport Systems}
                     International Civil Aviation Organization (ICAO) \cite{MRTD} has defined two different mechanisms to authenticate secure chips embedded in ePassports: active and passive authentication. In active authentication, the secure controller processes cryptographic information in the chip; in passive authentication, no computation is involved and the contents of a tamper-proof chip are read only by a verification device. Consequently, passive authentication is implemented on secure memory devices whereas active authentication requires a processor. Lately, a new type of authentication for EAC, called Chip Authentication has been proposed by European Union (EU) \cite{TREAC}.  Similar authentication was also proposed by some of the far east countries, but this paper details only the EU proposal.
                     
                     \subsection{Passive Authentication}
                     An ePassport's document security object \begin{math} {SO}_D \end{math} is digitally signed by issuing country at the time of personalization and the certificate is stored in secure chip. Hash of each data group (DG) is computed and stored in secure passport chip. In Passive Authentication (PA), the inspection system first verifies issuing country's signed security data object \cite{FIPS180} using public keys stored in the inspection system. If the signature matches, the hash of each data group is verified. By verifying hash of data groups inspection system infers if the data has been tampered. Certificate of document signer may be distributed to visited country in lieu of storing it on the chip. Typically, a visited country enters into an agreement with the issuing country to obtain the certificate and distribute it at different entry check points. Before checking the signature, validity of signed certificate is verified by checking Certificate Revocation List (CRL) for any updates. The revocation lists are regularly updated in a secure but mutually agreed storage area known as the Public Key Directory (PKD). The ePassport Public Key Infrastructure (PKI) symbols for passive authentication are defined in Section E of the appendix.

                     \subsubsection{Weaknesses}
                     Passive authentication does not prevent copying of chip data onto another chip, skimming, or unauthorized access to contents stored in the chip. The certificate of a document signer from an issuing country is stored on the secure chip. Reading a certificate from a secure chip to authenticate and verify a signature is not a good security practice, since it's quite possible that the certificate was revoked or it's invalid. The verification device must check the CRL while reading certificate from secure chip. For electronic passports the Certificate Revocation List (CRL) is stored in PKD.
                     
                     \subsubsection{Strengths}
                     An ePassport's security object has provisions to select hashing and signature algorithms. In case an algorithm in use turns obsolete, either because of weakness or otherwise, it can be switched to an alternate one. However, passports which have already been issued can not be substituted with the latest algorithm. The ICAO specifications for passive authentication have provisions for using larger key lengths which improve cryptographic security. The choice of a strong cryptographic algorithm for computing a hash of data structures improves on possible collisions of signature values. Passive authentication does not necessarily require the verification device to be online except to get the updated CRL from the PKD. This requires fewer infrastructure and can be performed in a remote location.
                     
                     \subsection{Active Authentication}
                     Active Authentication (AA) of electronic passports is performed using a unique cryptographic key pair \begin{math} KPu_{AA} \end{math} and \begin{math} KPr_{AA} \end{math}. The AA public key \begin{math} KPu_{AA} \end{math} is stored in Data Group 15 (DG15), one of 16 Data Groups (DGs), in the secure chip. The private key is stored in secure chip and never leaves the chip. Typically, an active authentication key pair is generated inside the secure chip; however, many system designers prefer creation of keys outside of the chip to improve personalization speed. The correctness of an AA key is verified by checking the signature of DG15, which is signed by passport signer's private key. To check  signature, signer's certificate should be retrieved from PKD or from the chip, and the CRL checked for updates. Visual inspection, electronic verification of the security object, and challenge response authentication using asymmetric key pair, determine if keys are read from a passport which has not been copied or cloned. The chip also stores signed MRZ information in the logical data structure (LDS). Active authentication is designed to detect if a passport chip has been replaced by a fake one or its contents have been copied to another chip. 

                     \subsubsection{Challenge Response}
                     The challenge response protocol for AA is outlined below.
                     
                     \begin{description}
                     	 \item[(a)] The authentication device checks for validity of the security object and retrieves the \begin{math}KPu_{AA}\end{math} either from the chip or PKD. 
                     
                     	\item[(b)] Before using the key in this cryptographic protocol, it checks for its validity by verifying the signature of the key. 
                     
                     	\item[(c)] The passport's chip contains the secret authentication private key \begin{math}KPr_{AA}\end{math} which is not accessible. The certificate and public key \begin{math}KPu_{AA}\end{math} are stored on the chip as well.
                     
                     	\item[(d)] The terminal generates a nonce and sends it to the secure chip as a challenge. The verification device may choose to send a date and time in addition to the nonce, which can optionally be stored in the chip upon signature. 
                     		\begin{equation}
                     			Challenge = R{VD} \rightarrow ePassport 
                     		\end{equation}
                     
                     	\item[(e)] The challenge, and at times an additional counter, is signed by the authentication private key. 
                     		\begin{equation}
                     			Response = Sign (RSA\footnote{The signature with message recovery is defined in ISO 9796-2. Other signature schemes like Digital Signature Algorithm (DSA per FIPS 186-3) and Elliptic Curve Digital Signature Algorithm (ECDSA) could also be used for the purpose. }){KPr_{AA}(R{VD}+C)}
                     		\end{equation}
                     
                     	\item[(f)] Verification device checks digital signature sent to it. 
                     
                     	\item[(g)] If the signature is verified, the secure chip is actively authenticated.
                     \end{description}

                     \subsubsection{Key Lifetime}
                     The authentication key pair is normally valid for the lifetime of an ePassport. Typically, ePassports are issued at least for five years or more. Therefore, the key lifetime of an active authentication key pair is five years. The strength of active authentication lies with the strength of the secure chip to securely hold the private key. Secure controllers with all the tamper-proofing mechanisms are strong key vaults. 
                     
                     \subsubsection{Weaknesses}
                     AA is designed to detect passports with cloned chips. It requires a secure chip with cryptographic capabilities, which may increase authentication time compared to PA. However, with faster processing and better algorithms this difference in time may not be noticeable. AA is usually performed in combination with BAC, which uses diversified keys that do not have high entropy. Use of AA with BAC overcomes the strongest weakness of BAC. In AA, the same key pair is used for every authentication session. There are no temporal keys for every new session of authentication. AA does not perform any type of external or terminal authentication. It assumes all terminals are trustworthy. This may not be an issue since, unlike second generation passports, first generation passports do not hold any private biometric data.
                     
                     \begin{figure}
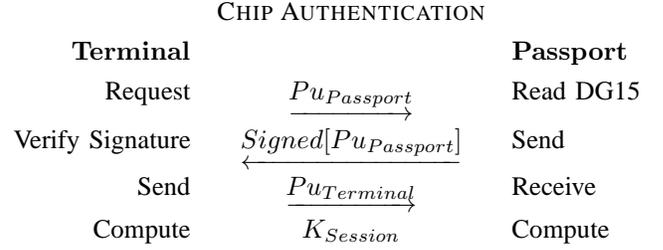

                     \newcounter{MYeqncnt}
                     \normalsize
                     \setcounter{MYeqncnt}{\value{equation}}
                     \setcounter{equation}{5}
                     \begin{IEEEeqnarray*}{r{'}C{'}l}
                     \text{} & \textsc{Chip Authentication} &  \text{} \\ 
                     \mathbf{Terminal} &  & \mathbf{Passport}  \\ 
                     \text{Request}  	& \underrightarrow{Pu_{Passport}} & \text{Read DG15} \\
                     \text{Verify Signature} & \underleftarrow{Signed[Pu_{Passport}]} & \text{Send} \\
                     \text{Send} & \underrightarrow{Pu_{Terminal}} & \text{Receive} \\
                     \text{Compute}  &     K_{Session}             & \text{Compute} \\
                     \end{IEEEeqnarray*}
                     \caption{Messaging in Chip Authentication}
                     \setcounter{equation}{\value{MYeqncnt}}
                     \hrulefill
                     \vspace*{6pt}
                     \label{fig:chipauth}	
                     \end{figure}
                     
                     \subsubsection{Strengths}
                     Introducing an asymmetric key pair allows signed trace and track information of visitors. An authenticated time stamp using the private key, which is stored only in the secure chip, allows system to be updated with trace information regarding entry or exit of a visitor. The private key is bound to the identity of person holding the ePassport. AA could improve privacy since each access to the passport can be logged in secure memory of the chip. The audit trail could be helpful in tracing and tracking an illicit request to access the passport. The AA public keys are signed and often stored on the chip; therefore, it does not require verification devices to be online.
                     
                     \subsection{Chip Authentication}
                     Chip authentication is used in second generation ePassports with BAC to improve security by introducing encryption of all messages exchanged between the inspection system and ePassport. In BAC with key diversification, the keys must be generated using holder-specific data which may not have high entropy. The active authentication and diversified keys are not unique to each and every session. Chip authentication improves these two authentication mechanisms by introducing keys which are unique to every chip and every session. Additionally, a message authentication code (MAC) is added to every message from the chip.  Like AA, it introduces asymmetric key pair, signed, with public key of the issuing country. In chip authentication, every chip has its own key pair assigned and stored in it. The public key is signed and stored in one of the public data groups and the private key is stored in secure memory of the chip \cite{Kugler}. See Figure~\ref{fig:chipauth}. 
                     
                     \subsubsection{Ephemeral Static Diffie-Hellman Key Exchange}
                     The terminal chooses an ephemeral key pair which is used to encrypt a single session of communication between the secure chip and the interface device. The Elliptic Curve Diffie-Hellman (ECDH) key agreement scheme is chosen to agree upon keys to encrypt data exchanged between chip and verification device \cite{Kugler}. Subsequently, a symmetric key, 3DES is used for encryption of all messages between the secure chip and terminal. This improves weaknesses of key diversification using static keys and PA challenge-response mechanisms. It also improves asymmetric key challenge response used in AA; the symmetric keys are used to encrypt messages. The following is a conceptual outline of chip authentication.
                                 
                     \begin{description}
                     	 \item[(a)] An elliptic curve and corresponding public curve point P is chosen for a given field and shared between the passport chip and verification device.
                     
                     	\item[(b)] The verification device generates a random number which is its private key. The passport's key, and private keys \begin{math}k_{p}\end{math} and \begin{math}k_{v}\end{math}, are stored in the chip. 
                     
                     	\item[(c)] Both generate public keys \begin{math}Q_{p}\end{math} and \begin{math}Q_{v}\end{math} and send them to each other.
                     
                     		\begin{displaymath} Q = 
                     			\begin{cases}
                     				Q_{p} = k_{p}P \\
                     				Q_{v} = k_{v}P  
                     			\end{cases}
                     		\end{displaymath}
                     
                     	\item[(d)] The chip and the verification device generate common shared data \begin{math}Q_{s}\end{math} using each other's public key.
                     
                     		\begin{displaymath} Q_{s} = 
                     			\begin{cases}
                     				k_{p}Q_{v} = k_{p}k_{v}P \\
                     				k_{v}Q_{p} = k_{v}k_{p}P 
                     			\end{cases}
                     		\end{displaymath}
                     
                     	\item[(e)] After the above step, \begin{math}Q_{s}\end{math} becomes the shared symmetric key for any encryption of any subsequent communication between the chip and the verification device. There are many different forms of the ECDH protocol, each with its merits. Menezes Qu Vanstone (MQV), another form of ECDH, has additional security mechanisms built in. An ePassport uses one of the forms of ECDH to perform the key agreement.
                     
                     \end{description}
                     
                     \subsubsection{Weaknesses} Chip authentication requires high-end processors which can perform Diffie-Hellman key exchange. Standard interfaces for Diffie-Hellman key exchange are available only in Java Card 2.2.x.
                     
                     \subsubsection{Strengths} Chip authentication covers all the weaknesses with the different schemes mentioned above. Along with message authentication as defined for BAC, chip authentication is the strongest known authentication mechanism.

                     \newcounter{MYtempeqncnt}
                     
                     \begin{figure}[h]
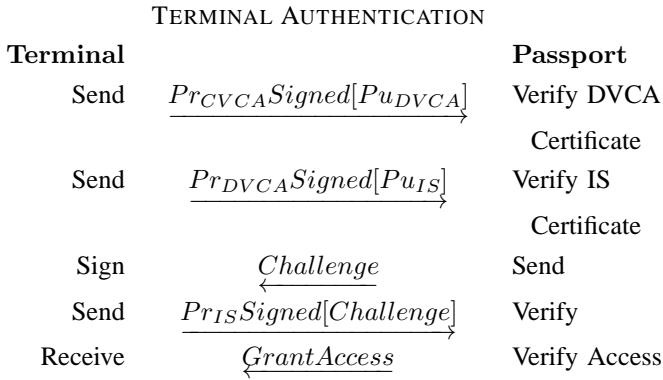

                     \normalsize
                     \setcounter{MYtempeqncnt}{\value{equation}}
                     \setcounter{equation}{5}
                     \begin{IEEEeqnarray*}{r{'}C{'}l}
                     \text{} & \textsc{Terminal Authentication} &  \text{} \\
                     \mathbf{Terminal} &  & \mathbf{Passport}  \\ 
                     \text{Send} & \underrightarrow{Pr_{CVCA}Signed[{Pu_{DVCA}}]} & \text{Verify DVCA} \\
                     \text{} &  & \hspace{.10in}\text{Certificate} \\
                     \text{Send} & \underrightarrow{Pr_{DVCA}Signed[{Pu_{IS}}]} & \text{Verify IS} \\
                     \text{} &  & \hspace{.10in}\text{Certificate} \\
                     \text{Sign} & \underleftarrow{Challenge} & \text{Send} \\
                     \text{Send} & \underrightarrow{Pr_{IS}Signed[Challenge]} & \text{Verify}\\
                     \text{Receive} & \underleftarrow{Grant Access} & \text{Verify Access} \\
                     \end{IEEEeqnarray*}
                     \caption{Messaging in Terminal Authentication}
                     \setcounter{equation}{\value{MYtempeqncnt}}
                     \hrulefill
                     \vspace*{6pt}
                     \label{fig:termauth}
                     \end{figure}
                     
                     \subsection{Terminal Authentication}
                     The second generation ePassport introduces concept of terminal rights and their authentication using asymmetric cryptography. In EAC \cite{TREAC}, holder's biometrics are stored on the chip. The holder's biometrics should not be released to terminals which do not have rights to read the information. A secure chip has no power source, so can not maintain time. Due to this limitation, the chip can not verify standard X.509 certificates. The secure chip in an electronic passport has limited access to the network, and therefore can not reliably update itself from the CRL. To overcome these limitations, a different type of certificate, known as a 'card/chip verifiable' certificate is used for external authentication.
                     
                     The terminal is authenticated if an asymmetric key pair is verified. The terminal stores private key of key pair and ePassport stores the public key, which is embedded with other information in a certificate, known as Card Verifiable Certificate (CVC). The CVC is securely stored to the ePassport at the time of personalization. The knowledge of private key in the terminal is verified by an asymmetric challenge response protocol. See Figure~\ref{fig:termauth}.
       
                     To authenticate the terminal, the ePassport sends a challenge to the verification device, which has access to the corresponding private key. The verification device attaches document number, challenge, and hash of the session-unique data and signs it with its private key. Either the RSA or ECDSA signature algorithm is used to sign and verify. The terminal contains its private key either in tamper-resistant memory or on the connected network from where it can be securely fetched. The chip contains 'trust anchor' or 'root certificate' which is used to verify signature received from the terminal. Built-in with authentication are access rights to read or update biometrics of the passport holder. Terminal authentication is preceded by internal chip authentication. As described above, chip authentication concludes with a mutually agreed symmetric key which envelops all the subsequent communication between ePassport and verification terminal.  All messages are encrypted and attached with MAC computed as per specifications of Chip Authentication.

                     \begin{figure}[h]
                     \centering
                     \includegraphics[height=1.5in,width=3.5in]{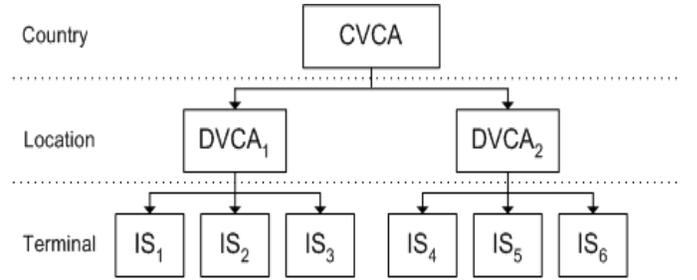}
                     \caption{PKI for Terminal Authentication}
                     \label {fig:PKIforTerminalAuthentication}
                     \end{figure}
       
                                  \subsubsection{PKI for Terminal Authentication} 
                     Figure~\ref{fig:PKIforTerminalAuthentication} illustrates the two-level PKI necessary for terminal authentication. Country Verifying CA (\begin{math}CVCA_{Issuer}\end{math}) issues certificate to document verifying CA (\begin{math}DVCA_{Issuer}\end{math}). The document verifying CA issues terminal certificates to each and every verification terminal deployed in the location. 
       
              \begin{figure}[h]
                     \centering
                     \includegraphics[height=1.5in,width=3.5in]{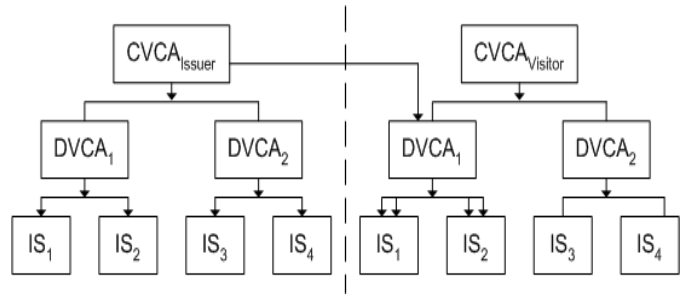}
                     \caption{PKI for Terminal Authentication with Cross Certification}
                     \label{fig:PKIforTerminalAuthenticationwithCrossCertification}
                     \end{figure}
       
                     Terminal verification extends beyond the boundaries of an issuing country. Cross certification allows foreign countries to verify the identity of the passport holder using biometrics. The \begin{math}CVCA_{Issuer}\end{math} certifies visited country's document verifying CA (\begin{math}DVCA_{Visit}\end{math}). The terminals at visited country now have signatures of the issuing country CA which can be verified by electronic passports from the issuing country. Figure~\ref{fig:PKIforTerminalAuthenticationwithCrossCertification} represents the PKI for cross certification.

                     \subsubsection{Access Rights}
                     EAC introduces access rights to verification terminals allowing only authorized terminals to read or modify certain data. The terminal's application has controlled access to different biometrics stored on the chip. Role-based access right mechanisms are implemented by encoded tables in card-verifiable terminal certificates. Such access right control mechanisms are used to limit or grant access to perform read or update operations. The card-verifiable certificates have been defined to contain privileges for different roles. Different roles include certificate authority, foreign country, and domestic inspection systems. Germany has chosen to implement such a system \cite{Kugler}.
                     
                     \subsubsection{Weaknesses}
                     Terminal authentication can not revoke a certificate once it has been issued to a country or an organization. Once issued, there is no known way to revoke a certificate using mechanisms like CRL and time. 
                     
                     \subsubsection{Strengths}
                     Terminal authentication adds terminal authentication and access rights to the facility of authentication, which is useful in controlling access to the biometrics stored inside the chip.

                     \section{Access Control in ePassport Systems}
                     
                     Access control using data stored in a secure chip is defined for various levels of classification by the issuing country. The U.S. had chosen Plain Access Control\footnote {Plain Access Control (PAC) is a term coined only in this paper by the author and may be viewed as special case of BAC. Per ICAO, BAC has mandatory document signing but the key diversification to read the data stored in chip is optional.} (PAC) as the primary method to control access. However, this was later changed to BAC after skimming and eavesdropping concerns were raised by different privacy groups. ICAO mandates storage of only a facial image in the chip; other biometrics, like fingerprints, are not stored in the chip, which limits the use of biometrics for access control. Mutual authentication of a secure chip and a verification device is optional. These limitations have been removed in EAC, which is better suited for biometrics-based access control at the verification point.\label{PassportPolicy}

                     \begin{table}[h]
                     \renewcommand{\arraystretch}{1.3}
                     \caption{Optional and Mandatory Security Mechanisms}
                     \label{table:20}
                     \centering
                     \begin{tabular}{l l } 
                     \hline 
                     Mechanism & Mandatory/Optional \\ [0.5ex] 
                     \hline\hline 
                     Passive Authentication & ICAO Mandatory \\ 
                     BAC & ICAO Optional, EU Mandatory \\
                     Active Authentication & Issuing Country \\ 
                     EAC & EU Mandatory for biometrics \\
                     \hline 
                     \end{tabular}
       	        \label{table:optionalmandatory}
                     \end{table}
       
                                  \subsection{Plain Access Control }
                     
                     PAC is a special case of BAC,  in which key diversification is not required to read the data. This scheme of access control allows any reader to read data from a chip. The secure data object is hashed, signed, and stored in the LDS. Authenticity of a passport is verified by checking the authenticity of data and its signature via a signed object. According to ISO 14443, read range is limited to 10 cm; however, the data can be skimmed using modified readers that achieve a range greater than 10 cm. Therefore, PAC does not counter skimming and eavesdropping types of attacks. PAC may be secure for contact only communication; however, contactless communication introduces risks and mere signature verification is inadequate security.  See Table~\ref{table:optionalmandatory}.

                     \subsubsection{Security Weaknesses}
                     
                     \paragraph{Confidentiality}
                     The data is unencrypted and can be read by any reader. There is no confidentiality of data and this scheme allows any reader to read the data \cite{Schneier}.
                     
                     \paragraph{Authentication}
                     Any reader can read the data encoded on the chip. There is no security mechanism to authenticate the reader, passport holder, or the secure chip.
                     
                     \paragraph{Data Cloning}
                     The data can be easily read and written to another chip without any modification.

                     \subsection{Basic Access Control}
                     Per ICAO, BAC is an optional but recommended way to achieve interoperability of ePassport-based border control between countries since the European Union mandated diversified key authentication. The diversified key is generated, using MRZ data, and used to mutually authenticate the passport and inspection devices. Since common knowledge is required to generate the diversified keys, this scheme implicitly authenticates the inspection system; however, mutual authentication is not strong. In BAC, data between inspection or verification reader is not encrypted using the session keys. An ePassport could be  authenticated using either active or passive authentication.   
                     
                     The mandatory and optional stages of BAC are listed below. 
                     \begin{enumerate}
                         \item Key Diversification [Optional]: BAC diversified key authentication and opening of an unencrypted communication channel is an optional but suggested method to authenticate the ePassport and inspection device.
                     	\item Passive Authentication [Mandatory]: Passive authentication and checking of the signature is a mandatory step in BAC. Passive authentication verifies authenticity of the data only.
                     	\item Active Authentication [Optional]: Active Authentication, as described above, is an optional step in BAC. 
                     \end{enumerate}
                     
                     \subsubsection{Key Diversification}
                     The fixed seed key is diversified using the passport number, date of birth, and expiration date of the document. Key diversification creates a unique key for each passport; however, keys are diversified using known data which is not random and therefore has low entropy. A true random number can not be known to both the inspection system and the secure chip without doing one of the following.
                     \begin{itemize}
                        \item Use a back-end system to store a known secret random number. The same number can be stored in the card.
                     	\item Exchange the random number as part of a setup message. However, this must be passed in the clear, which weakens security.
                     	\item Use a known but secret coding algorithm to generate the same codes. A known algorithm could be used to generate a pseudo random number. However, this is not a good approach since a secret black box algorithm could be broken. 
                     \end{itemize} 
                     
                     \subsubsection{Session Key Derivation}
                     BAC uses two keys, one for encryption and another for calculating the message authentication code being exchanged between reader and secure chip. Sessions keys are generated using a seed key derived using the data read from the MRZ. 
                     
                     \begin{enumerate}
                     	\item The passport contains \begin{math}K_{seed}, MRZ Data, C_{Enc} \end{math} and \begin{math}C_{MAC}\end{math}. The session keys are derived from seed key and personal data stored in MRZ.
                     \begin{displaymath}
                     	\begin{array}{|cc|}
                      		K_{seed} & C_{MAC}\\
                      		MRZ Data & C_{ENC}
                     	\end{array}
                     \end{displaymath}
                     
                     	\item The seed key \begin{math}K_{seed} \end{math} is diversified using data stored in the MRZ.
                     		\begin{displaymath}
                     			MRZ Data \otimes K_{seed} \rightarrow K_{div}  
                     		\end{displaymath}
                     
                     	\item The \begin{math}K_{div} \end{math}  is concatenated with 32-bit counters. One for encryption and the other for MAC.
                     		\begin{displaymath} D = 
                     			\begin{cases}
                         			K _{Enc} \Leftarrow K_{div} || C_{Enc},\\ 
                     				K _{MAC} \Leftarrow K_{div} || C_{MAC} 
                     			\end{cases}
                     		\end{displaymath}
                     
                     	\item Next, a message digest is generated on key set D. SHA-1 is calculated on both the keys defined in key set D, now called H. This is a 160-bit or a 20-byte long number.
                     		\begin{displaymath} H = 
                     			\begin{cases}
                         			H _{Enc} \Leftarrow SHA ( K_{Enc} ) ,\\ 
                     				H _{MAC} \Leftarrow SHA ( K_{MAC} ) 
                     			\end{cases}
                     		\end{displaymath}
                     
                     	\item The first 56 bits of H constitutes K(a) and checksum is calculated per the DES algorithm. The computed checksum is appended to the key to make it 64 bits long.
                     		\begin{displaymath} H \rightarrow K_{a}
                     		\end{displaymath}
                     
                     	\item The next 56 bits from the 64th bit of H constitutes K(b) and a checksum is calculated per the DES algorithm. The computed checksum is appended to the key to make it 64 bits long.
                     		\begin{displaymath} H \rightarrow K_{b}
                     		\end{displaymath}
                     
                     	\item The last 20 bits or four bytes of H are discarded with no affect.
                     
                     	\item The challenge response messaging occurs to verify the passport as per ISO 11770-2 using 3DES in block-cipher mode.
                     
                     	\item Random numbers (nonces) generated by the passport and the reader are of eight bytes long.
                     \end{enumerate}
                     
                     The keys generated are valid from the start to the end of communication between the verification device and ePassport. The lifetime of the document signing key is equal to the longest time for which the passport is valid plus time for which the key has been used to sign other passports. The document-signing key is erased once the key has expired. Countries choose the frequency at which a new document-signing certificate is issued.
                     
                     \subsubsection{Challenge Response}
                     The message sequence for challenge response is summarized below. The nonce is eight bytes or 64 bits long.
                     
                     \begin{enumerate}
                     
                     	\item The verification device generates a random number 								\begin{math}R_{v} \end{math} and encrypts with the triple DES keys generated.
                     	\begin{displaymath} 
                     		M_{vp} = 3DES(K_{ab}(R_{v})) 
                     	\end{displaymath}
                     
                     	\item The passport decrypts it and verifies if the random number matches.
                     		\begin{math}R_{v} \end{math}encrypts with the triple DES keys generated.
                     			\begin{displaymath} 
                     				R_{v} = 3DES^{-1}(K_{ab}(R_{v}))  	
                     			\end{displaymath}
                     
                     	\item Passport generates a random number \begin{math}R_{p} \end{math} encrypts with the triple DES keys.
                     	\begin{displaymath} 
                     		M_{pv} = 3DES(K_{ab}(R_{p})) 
                     	\end{displaymath}
                     
                     	\item The verification device decrypts the challenge and verifies if the random number matches.
                     		\begin{math}R_{p} \end{math}encrypts with the triple DES keys generated.
                     			\begin{displaymath} 
                     				R_{p} = 3DES^{-1}(K_{ab}(R_{p}))  	
                     			\end{displaymath}
                     \end{enumerate}
                     
                     \subsubsection{Message Authentication}
                     Every message between passport and reader can be appended with a Message Authentication Code (MAC). The MAC is calculated per the ISO 9797-1 MAC algorithm. The MAC is calculated using the DES algorithm with Cipher Block Chaining (CBC). For all blocks except \begin{math}B_{N-1}\end{math}, \begin{math}K_{a}\end{math} is used for encryption. For every block, an incremental counter is appended to the message to make its MAC unique. This technique works well against message replay attacks, which are common in wireless communication. Note that the next to last block is encrypted using \begin{math}K_{b}\end{math} instead of \begin{math}K_{a}\end{math}.
                     
                     \subsubsection{Security Weaknesses}
                     
                     \paragraph{Key Generation}
                     Both the encryption and the MAC key are generated from the same seed key. Although there are additional counters attached to generate the encryption and MAC keys, the compromise of one single key \begin{math} K_{seed} \end{math} could be sufficient to break the encryption as well as the message authentication code. Note that the counter is only an additional number incremented by one. To overcome this threat, the counters can be easily replaced by a random number which can be either encrypted or attached to the message.
                     
                     \paragraph{Key Diversification}
                     The seed key \begin{math} K_{seed} \end{math} is diversified using the passport holder's name and date of birth, or the keys are diversified using the passport's information which can be easily read from the data page of the passport. The entropy of the diversification information is limited and the diversification information is available in the data page of the passport. This data in most cases is encoded in the machine-readable zone. Despite the use of the SHA 200 hashing algorithm, the key generated may not be truly random. 
                     
                     \paragraph{Cryptographic Algorithm}
                     The choice of cryptographic algorithm has been DES, a well known and proven standard for symmetric encryption/decryption of data. However, with Rijndael's algorithm already announced as the AES, the choice of DES does not seem appropriate. AES has already been designed into some networking and banking protocols. AES's stronger substitution permutation network structure and the fact that it can be implemented well on a medium- or small-sized chip, like the one used for passports, makes it a better candidate than DES. The lack of adoption of AES in the ePassport standards should probably be revisited sometime.
                     
                     \paragraph{Chip and Data Cloning}
                     The secure chip can not be easily protected against a clone attack. A cloned chip is assumed to have copied all data bit-wise and replicated to similar silicon. Detection of such clones is a challenge which industry is presently facing. The anti-cloning solution may lie beyond the secure chip is some special type of printing or material that is added to passports.
                     
              \subsection{Extended Access Control}
                     Following rollout of U.S. passports, the EU came up with its set of passport identification standards and security protocols \cite{TREAC}  \cite{Nguyen}. EAC is designed for a second generation of passports which involves storing biometrics of the holder. The common biometrics used are either finger prints or iris images. Terminal Authentication is being implemented by EU countries with some country-specific extensions. EAC keys are exchanged bilaterally between the issuing and visited state. The signed certificates must be available to the verification device for reading biometric information stored on the secure chip. 
                     
                     EAC replaces active authentication by chip authentication. As per \cite{VKrishnan1}, an EAC mutual authentication session consists of the following stages with mandatory and optional parts. 
                     \begin{enumerate}
                        \item BAC [Mandatory for all Passports]: Establishes a secure channel with diversified keys. 
                     	\item Chip Authentication [Mandatory for Second Generation Passports]: EAC replaces active authentication by mandatory chip authentication.
                     	\item Passive Authentication [First Generation Passports]: Passive authentication is performed as per ICAO 9303 specifications. 
                     	\item Terminal Authentication [Mandatory for Second Generation Passports]: EAC adds terminal authentication to mutual authentication. All terminals are not trusted by the passport, as is the case for first generation passports.
                     \end{enumerate}
                     
                     EAC message exchange consists of two stages: chip authentication followed by terminal authentication. The chip is ``deduced'' to be verified at the beginning of the second stage since it is able to encrypt and decrypt messages using session keys negotiated in the first stage. The session keys are established in the chip authentication stage. The first stage consists of an Elliptic Curve Diffie-Hellman or Diffie-Hellman key exchange. The terminal authentication stage consists of signature verification using either RSA or ECDSA.
                     
                     \subsubsection{Security Weaknesses}
                     \paragraph{Card Verifiable Certificates}
                     EAC involves use of card-verifiable certificates which are not standard. Unlike X.509 and PGP certificates, such non-standard certificates are not widely deployed. There have been limited implementations \cite{X509Cert} of verification of X.509 certificates on smart card chips. A card-verifiable certificate can not be revoked or its validity checked with time since cards do not have a clock. This weakness limits the effectiveness of security mechanisms based on card-verifiable certificates.
                     
                     \paragraph{Loss of Verification Terminal}
                     In extended authentication schemes, a private key is stored in the terminal; therefore any theft or loss of a verification terminal could jeopardize the security of all passports using the corresponding public key. This could possibly mean a breach of security; however, to overcome this weakness, tamper-resistance mechanisms should be deployed to secure the storage of keys and sensitive information.

                     \section{Biometrics}
                     Facial images and fingerprints are the most useful biometrics for ePassports \cite{Nusken}. Biometrics add another dimension to the authentication of ePassports by allowing checks for ``what you have'' and ``what you are'' at the same time. Biometrics can help to move from manned border crossing stations to unmanned border verification stations. The automated verification of credentials was first implemented by Malaysia and followed by Australia. Brazil has decided to roll out its ePassports with all ten fingerprints and a facial image encoded on the chip. 
                     
                     Many countries use biometrics for law enforcement, using a one-to-many search-and-match to identify criminals. Since criminals often try to cross country  borders, having similar mechanisms such as FIPS 201 or Personal Identity Verification (PIV) is useful for ePassports. A one-to-many comparison requires additional storage and computing for quick results. This can be achieved only by letting the minutiae out of the secure chip and matching them to the database of foul fingerprints. EAC grants or denies access based on biometric information stored in the secure chip. Just like electronic signatures, a biometrics template is unique to the passport holder, and therefore raises different privacy concerns about the distribution of template information. Storing biometrics securely in the secure chip with access control is the best way to maximize security and privacy concerns. As per ICAO, the facial image of a passport holder is not sensitive or secret biometric data. A facial image is used in BAC for personal identification and therefore it is not encrypted. Table~\ref{ref:biometrics} shows provisions available in electronic passports for additional biometrics like an iris or retinal image and visual marks.

                     \begin{table}[h]
                     \renewcommand{\arraystretch}{1.3}
                     \caption{ePassport Biometrics}
                     \label{table:5}
                     \centering
                     \begin{tabular}{ l l } 
                     \hline 
                     Name & Classification \\ [0.5ex] 
                     \hline\hline 
                     Facial Image & Public \\
                     Encoded Finger & Private \\
                     Encoded Iris/Retina & Private \\
                     Visual Mark & Private \\
                     Signature & Private \\
                     \hline 
                     \end{tabular}
                     \label{ref:biometrics}
                     \end{table}

                     \subsection{Electronic Verification}
                     Electronic verification is possible only if the passport holder is present and the verification station has the necessary devices to capture the biometric data of the passport holder. If the biometrics data is encrypted, it must be decrypted after generating the diversified keys which are calculated only after reading the MRZ. The biometrics of the passport holder are captured using a capturing station and biometrics sensor and matched against those stored on the secure chip. ICAO specifications do not mandate a comparison of biometrics on the secure chip, commonly know as ``match on chip'' Since the match is typically done on the verification device, to secure the verification the biometrics should be encrypted since they will travel wirelessly.
                     
                     Besides the risk of losing minutiae to an illegitimate terminal, the verification of fingerprints at an unmanned station introduces the 'gelatin finger' or 'prosthetic finger' threat. The threat of introducing a prosthetic or gelatin finger with ridges as per stolen images of an individual can not be easily mitigated. The author can propose reading more than one finger to reduce the chances of such a threat succeeding. Enrolling all ten fingers on the chip and randomly selecting the finger to verify helps improve security. The security against this threat also improves if more than one fingerprint of the individual is verified. Extending this logic, if both the finger and facial biometrics are stored on the chip and electronically verified at the unmanned station, the gelatin finger threat is mitigated. Experts have argued that the facial images could also be easily found and replicated to fool the facial recognition systems; however, security with ``dual checks'' should be an acceptable level of security.
                     
                     \subsection{Terminal Verification}
                     All terminals are usually not granted access to read the biometric templates stored on the card. Some countries like Germany have designed their systems to validate the access rights of terminals to read the template before exposing templates to them. Terminal authentication is performed and the access bits decoded to check for the access rights.

                     \section{Protection Profiles}
       
                     Common criteria profiles \cite{PPBAC} \cite{PPEAC} have been defined for each type of ePassport access control. Most of the hardware, embedded software, manufacturing process and issuance, and handling systems are evaluated to meet high degrees of assurances. The two different protection profiles cover threats as perceived by the developers of such profiles. The protection profiles may be comprehensive but may not be complete. An example of such a threat is a side-channel attack. Protection against such an attack has been built in most of the secure controllers.

                     \section{Conclusions}
                     Since ePassports are being issued in place of traditional paper passports, which do not have secure RF-enabled contactless chips, the security of ePassports must be better than paper-based passports, which are not vulnerable to skimming, eavesdropping, or tracking attacks. This is an unwritten and probably unrealized expectation from the issuing countries, passport holders, and society. The world is seeing the first few generations of RFID passports. The convenience and security of contactless access control transactions are here to stay. The second generation passport with biometric access control will be more prevalent in the coming years. The future of passports may shift from single-chip electronic RF to multi-chip modules with a combined data storage capacity from multiple chips. Future ePassports may look like a secured solid disk with onboard sensors and limited or no printed information. This would be a true generation shift from paper electronic to fully electronic passports. Such a passport may have its own sensors on board to validate its holder.  Some other interesting security features like RF-DNA \cite{DeJean} \cite{Yuqun} can be used as certificates of authenticity in lieu of electronic certificates.  These methods rely on physical creation of fingerprints which would identify each passport.  See Table~\ref{ref:generation}.
                     
                     \begin{table}[h]
                     \renewcommand{\arraystretch}{1.3}
                     \caption{Generations of Passport}
                     \label{table:8}
                     \centering
                     \begin{tabular}{ l  l } 
                     \hline 
                     Generation & Type \\ [0.5ex] 
                     \hline\hline 
                     Generation 0 & MRZ \\ 
                     Generation 1 & BAC + PA, EA \\
                     Generation 2 & EAC + TA, CA \\
                     Generation 3 & eVisas \\
                     Generation 4 & Advanced Electronics \\
                     \hline 
                     \end{tabular}
                     \label{ref:generation}
                     \end{table}

                     
                     %

                     \appendix
                     
                     \subsection{Construction}

                     \subsubsection{Inlay}
                     The tamper-proof chip embedded in a passport is connected to a coil and made into an inlay which is inserted on either the cover or first page of a machine-readable travel document (MRTD). The term MRTD refers to ePassports as well as other travel authorization documents including visas and permits for entry. The inlay is usually composed of different layers of synthetic polymers which protect the coil, chip, and more importantly the interconnect. The most important of all components is a secure RF-enabled tamper-proof chip which allows reading of data while securely storing it. An ePassport's physical construction is required to last for at least 10 years. The physical construction should be strong enough to survive the pressure and temperature that it could be subjected to in those years \cite{VKrishnan1}.                

                     The construction of a passport consists of layers of laminated poly-carbonate or polyester (PET/PETG) or some other material which is a combination of similar materials \cite{Smartrac}. The inlay is usually 300-480 \begin{math}\mu\end{math}m thick, depending on the thickness of the silicon. The dimensions of the inlay are about 600 x 400 mm in length and width. The chip is made into a module before bonding the antenna to the module. Often, layers of different materials are used to strengthen the composition of the inlay. An inlay goes through numerous test cycles, including the ISO 10373 physical, temperature, magnetic, electric, and chemical tests. Besides this, an inlay must go through another important test known as the high-pressure impact or ``stamping'' test. Since passports are often stamped, they need to survive the stamping impact stress.

                     \subsubsection{Antenna}
                     The antenna is typically made of copper or some other alloy of copper to construct a coil with the right attenuation and RF characteristics. The antenna dimensions of a passport are specified in ISO 7810 ID-1. The size of such an antenna is between 85.60 and 53.98 mm (3.370 to 2.125 in). The number of turns in the coil is not specified but left to the antenna designer. The same goes for material used to manufacture the antenna. The electrical and magnetic characteristics of the magnetic field are specified in ISO 14443.

                    \begin{figure}
                     \centering
                     \fbox{\includegraphics[width=2.5in]{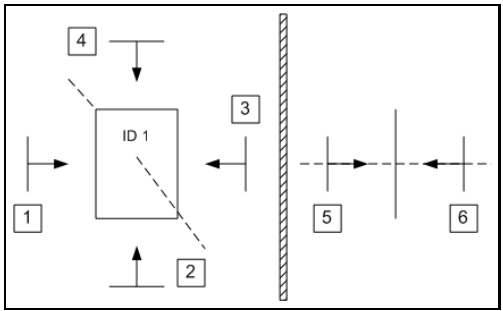}}
                     \caption{Directions for Measuring Attenuation}
                     \label{fig:field}
                     \end{figure}
             
                     \subsubsection{RF Shield}
                     An RF shield is a mesh of electro-magnetically opaque material, usually put on the cover of a passport to avoid skimming and eavesdropping \cite{9303}. The shield, also known as a ``Faraday Cage'',  is often in the form of a pouch or cover on the passport \cite{IEEE299}. The shield offers -40 to -60 dB attenuation in the frequency band around 13.56 MHz to limit skimming or eavesdropping. This level of attenuation is perceived to block the communicating signals between a passport and interrogating readers. An RF shield is not fully effective in blocking communication between a reader and a passport. A reader with strong signal can be designed that could nullify attenuation provided by an RF shield when the ePassport is open and the shield is not effective.  Shield is less effective if it is placed in the cover page and the cover page is opened.  The shield provides some safety but not the surety of blocking communication between a passport and a rogue-interrogating reader. It should not be relied upon as the sole mechanism to avoid such attacks. The exact methods of measuring the attenuation of electro magnetic shielding is defined in \cite{IEEE299}. For a geometry of the ID1 format, attenuation should be calculated at all six sides and preferably at regular angular intervals of inclination. The different directions of measurement are illustrated in Figure~\ref{fig:field}.

                     
                     \subsubsection{The Secure Chip} 
                     The secure chip module is usually smaller than 20 mm\begin{math}^{2}\end{math}. Of six pins of a secure chip, only two are connected to the antenna. Other pads that are not used must be secured against probing attacks. The contactless air interface could be either Type A or B. Per ICAO, the minimum size of available persistent non-volatile memory on the chip should not be less than 30 kB. Chips with 30 kB of available data space, are sufficient only for storing a minimal set of mandatory biometrics and can be used only for first-generation ePassports. Higher capacity chips are required for countries intending to store visas and travel information on the passport. The writable information on a chip includes data groups DG17, DG18, and DG19. The significance of each data group is explained in Section \ref{Appendix:LDS} below. In such cases, the minimum size of persistent non-volatile memory must be greater than 256 kB. The chip should be capable of generating random numbers and performing at least DES/3DES cryptographic operations like encryption, hashing, and signing for BAC. The chip must be capable of performing asymmetric cryptographic operations like Elliptic Curve Diffie-Hellman key exchange for EAC.
                     
                     \subsubsection{Radio Frequency}
                     All ePassports are specified to work only at 13.56 MHz for the advantages of high-frequency communication over low frequency. The 13.56 MHz communication band is more immune to processor noise, noise from the earth's field, and the surrounding environment, which is necessary for stable and consistent RF communication. The RF communication is based on ISO 14443 standards developed for near-field inductively coupled systems. ICAO defines communication to be either Type A or Type B. ICAO has additional test standards for stronger adherence to field strengths and modulation indices. Also defined are strong test criteria for its qualifications.

                     
       
                     

                     \subsection{Logical Data Structure}\label{Appendix:LDS}
                     To achieve interoperability, the LDS of electronic passports is defined by an ICAO standard. The data structure has various mandatory or optional components. Data groups DG 1-16 are written by the issuing country; data groups DG 17-19 are written by a receiving country.


                     \subsubsection{Mandatory Issuing State or Organization Data}
                     The mandatory encoded LDS is shown in Table~\ref{table:1}. This includes two data groups DG1 and DG2.
                     
                     \begin{table}[h]
                     \caption{Mandatory Logical Data} 
                     \centering 
                     \begin{tabular}{c c l } 
                     \hline 
                     Location & Group & Type \\ [0.5ex] 
                     \hline\hline 
                     MRZ + Chip & DG1 & Holder and Passport Info. \\ 
                     Data Page + Chip & DG2 & Encoded Face of Holder \\
                     \hline 

                     \end{tabular}
                     \label{table:1} 
                     \end{table}

                     \begin{table}[h]
                     \caption{Issuing State or Organization Logical Data} 
                     \centering 
                     \begin{tabular}{ c l } 
                     \hline 
                     Name & Data \\ [0.5ex] 
                     \hline\hline 
                     DG3-DG7 & Holder's Biometrics\\    
                     DG8 & Data Feature(s) \\
                     DG9-DG14 & Structure Feature(s) \\
                     DG10 & Substance Feature(s) \\
                     DG11-DG14 & Additional Info / RFU \\
                     DG15 & Active Authentication Key Info \\
                     DG16 & Person to Notify \\
                     \hline 
                     \end{tabular}
                     \label{table:2} 
                     \end{table}
                     

             \subsubsection{Optional Logical Data}
                     The optional LDS which can be encoded is shown in Table~\ref{table:2}.

                     \subsection{Electronic Passport Public Key Infrastructure}\label{Appendix:PKI}
                    An  ePassport requires a PKI for every issuing country and a cross state (central) repository to exchange any updates. A PKI is required even for BAC and passive authentication of ePassports. Every issuing country has its own Country Signing Certificate Authority (CSCA) and Country Verifying Certificate Authority (CVCA). A CSCA's public signature key is denoted as \begin{math}KPu_{CSCA}\end{math} and its private signature key is  \begin{math}KPr_{CSCA}\end{math}. A CVCA's public signature key is denoted as \begin{math}KPu_{CVCA}\end{math} and its private signature key is  \begin{math}KPr_{CVCA}\end{math}. 
                     
                     A CSCA issues certificates to the document signer which are used to sign the LDS. The document signer's public key is denoted as \begin{math}KPu_{DS}\end{math} and the private key is \begin{math}KPr_{DS}\end{math}. A CVCA issues certificates to a Document Verifier (DV), which may or may not be the same entity as the document signer. For active authentication, an additional key pair is added. This key pair is only used in a challenge response to actively authenticate the ePassport chip. The public active authentication key is denoted as \begin{math}KPu_{AA}\end{math} and the private key is denoted as \begin{math}KPr_{AA}\end{math}. The authenticity of an ePassports is verified by the document verifier. The issuing country can revoke certificates and have a need to update the CRL and set of document verification certificates. These certificates are normally stored in a common shared and secure repository, which is regularly scanned for updates.

                     \end{twocolumn}


\end{document}